Zentropy Theory for Quantitative Prediction of Emergent Behaviors through Symmetry-Breaking Configurations

Zi-Kui Liu

Department of Materials Science and Engineering, The Pennsylvania State University, University Park, Pennsylvania 16802, USA

**Abstract:**

Based on statistical mechanics, the partition function of a system is a sum of partition functions of all configurations that the system embraces. The key challenge is to find those configurations and obtain their properties. While the usual approach is to define those configurations in terms of the basins on the energy landscape, the zentropy theory developed by the author's group builds the configuration ensembles starting from the ground-state configuration at zero K based on the density functional theory (DFT) plus the symmetry breaking non-ground-state configurations through its internal degrees of freedom.  DFT is the de facto approach for predicting self-consistent-field electronic structures of ground-state configurations of complex atoms, molecules, and solids.  This capability is greatly enabled by the generalized gradient approximation (GGA) for exchange-correlation interactions with an important set of exchange-correlation functionals developed by John Perdew and his collaborators in last several decades including the latest strongly constrained and appropriately normed (SCAN) meta-GGA for more accurate ground-state energy and self-interaction correction by symmetry breaking configurations.  With the free energies of all configurations predicted by the DFT-based calculations, the zentropy theory postulates that the total entropy of the system includes both the statistical configurational entropy among all configurations and the entropy within each configuration, resulting in that the partition function of each configuration must be evaluated using its free energy rather than commonly used total energy. Consequently, the calculations from the ground-state configuration alone should not be quantitatively compared with experimental observations in general. It is articulated that phonon properties of all configurations can be accurately calculated by quasiharmonic approximations, and the emergent behaviors and anharmonicity of a system originate primarily from the statistical competition among all configurations.

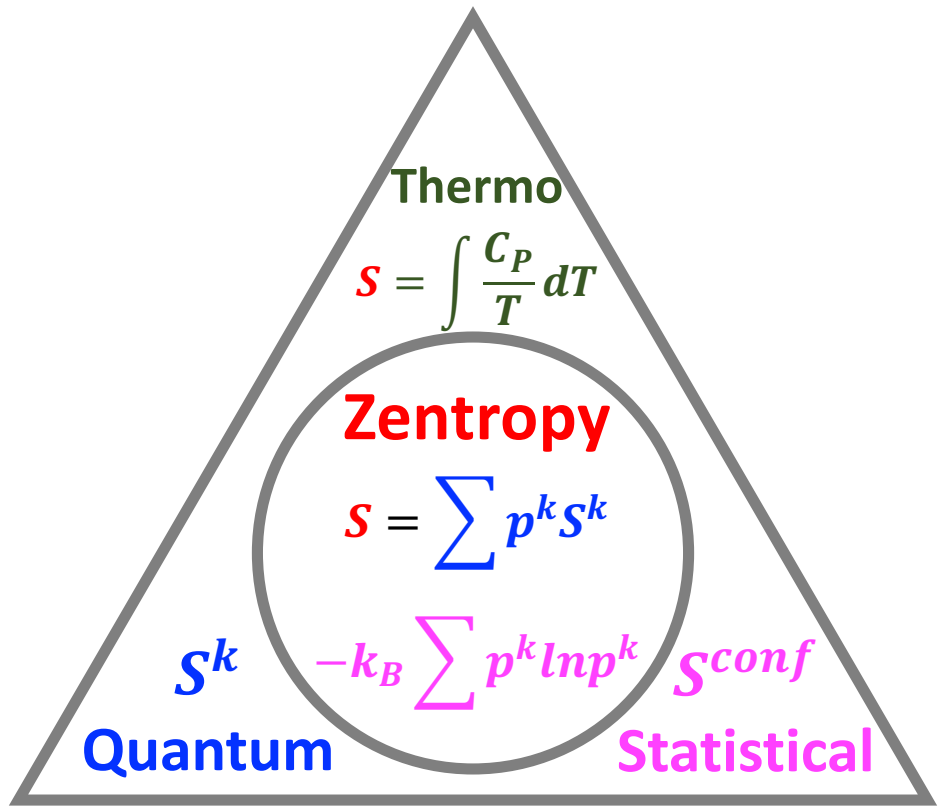

*Schematic representation of the zentropy theory for total entropy S with contributions from each configuration k via DFT-based calculations ($S^k$) and configurational entropy among configurations ($S^{conf} = -k_B \sum p^k \ln p^k$) with $p^k$ for probability of each k, which is equal to the entropy from the integration of heat capacity ($C_P$) from experiments.*

# 1   Introduction

The discovery of quantum mechanics (QM) is one of the most significant events that dramatically enhanced our understanding of nature. The numerical solutions of the QM Schrödinger equation [1,2] of complex atoms, molecules, and solids are offered by the density functional theory (DFT) [3,4] and have enabled the scientific community to predict unknowns and accumulations of new data and knowledge from computation. DFT articulates that for a given system, there exists a ground-state configuration that its energy is at its minimum value with a universal functional of the interacting electron gas density [3]. This unique ground-state electron density is obtained by explicitly separating the independent-electron kinetic energy and long-range Coulomb interaction energy and replacing the many-body electron problem using independent valence electrons with an exchange-correlation functional of the electron density and an associated exchange-correlation energy [4], i.e., *coarse graining of electrons*.

On the other hand, a system at finite temperature is represented by a statistical mixture of various configurations from the macroscopic view of the system, i.e., a top-down view in terms of *coarse graining of configurations* in contrast to the bottom-up view of the ground-state configuration in DFT. In introducing statistical mechanics (SM) in 1901, Gibbs [5] considered "a great number of independent systems of the same nature, but differing in the configurations and velocities which they have at a given instant, and differing not merely infinitesimally, but it may be so as to embrace every conceivable combination of configuration and velocities". It is thus evident that the macroscopic structure and properties of a system are the statistical average of those configurations, while the fluctuations of those configuration with respect to external conditions result in change of macroscopic properties of the system.

Since DFT was invented in 1960s, the DFT and SM communities have been largely separated from each other with the former focusing on the ground-state configurations of various systems and the latter based on Gibbs formalism. Efforts have been made to bridge the gaps between them through bottom-up approaches by considering the thermal electronic and phonon distributions of the ground-state configurations or using effective Hamiltonian fitted to DFT-based calculations and/or experimental observations followed by molecular dynamics (MD) or Monte Carlo (MC) simulations. There are also limited work on *ab initio* molecular dynamic (AIMD) and quantum Monte Carlo (QMC) simulations. However, quantitative agreement between predictions and experiments is rare in the literature due to the intrinsic limitations of existing approaches as discussed in our recent publication [6], i.e., the simultaneous considerations of all internal degrees of freedom and the ergodicity of independent configurations.

The author's group started to explore an approach to integrate DFT and SM in efforts to predict the temperature-pressure (T-P) phase diagram of Ce in 2008 [7] and consider the necessary ergodicity of configurations in Ce in 2009 [8] that enabled the accurate prediction of its T-P and temperature-volume (T-V) phase diagrams due to magnetic spin configurations. Without prior models imposed on the magnetic transition, thus free from fitting model parameters, the critical point and the positive divergency of thermal expansion at the critical point were predicted. Those predictions are in quantitative agreement with available experimental observations reported in the literature. In 2010 [9], the author's group made the model- and parameter-free prediction of the T-P and T-V phases diagrams of $Fe_3Pt$, its negative thermal expansion (NTE), i.e., INVAR effect, the NTE temperature range as a function of pressure, its critical point in the T-P and T-V phase diagrams, and the negative divergency of NTE at the critical point. Those predictions are again in quantitative agreement with available experimental observations in the literature. In 2021, the term "zentropy" was suggested to represent the approach [10], followed by its application to strongly correlated $YNiO_3$ in 2022 [6] and ferroelectric $PbTiO_3$ in 2023 [11].

In this short review for the special issue of John Perdew Festschrift of 80th birthday, the zentropy theory is presented along with some predicted results reported by the author's group. It is noted that Wentzcovitch's group [12–19] and Allan's group [20,21] have independently worked with similar approaches as discussed recently by the author [22]. The rest of the paper is organized as follows with Section 2 on DFT-based calculations, Section 3 on zentropy theory, Section 4 on general discussions of applications, and the summary at the end.

## 2 DFT-based Calculations

Building on the local density approximation (LDA), Perdew and co-workers developed the generalized gradient approximation (GGA) [23–25], in which the exchange-correlation energy is treated as a function of both the local electron density and its gradient, resulting in more accurate predictions of electronic structure and the energy of ground-state configurations. Over the years, they and the community have developed various GGA functionals to improve the prediction of DFT-based calculations, including the latest strongly constrained and appropriately normed (SCAN) meta-GGA [26,27] with quantitatively correct ground-state results by symmetry breaking for some systems regarded as strongly correlated [28,29] and the $r^2$SCAN with both improved accuracy and numerical stability and efficiency [27,29–32]. The key discovery in the SCAN meta-GGA is that strong correlations within a symmetry-unbroken ground-state wavefunction can show up in approximate DFT as symmetry-broken spin densities or total densities due to soft modes of fluctuations such as spin-density or charge-density waves at nonzero wavevector [28]. Consequently, an approximate density functional for exchange and correlation with symmetry breaking, though less accurate than an exact functional, can be more revealing with its utility demonstrated for a number of cases [28,29,32].

The continuously improved GGA functionals with more and more accurate ground-state properties have resulted in ever-growing massive digital databases of properties predicted using high-performance computers and potentials developed by Perdew, his collaborators, and the community. Many databases are linked with the Open Databases Integration for Materials Design (OPTIMADE) consortium [33] with a universal application programming interface and an extensive list of database providers [34], including the Material-Property-Descriptor Database (MPDD) from the author's group [35,36]. In predicting the diffusion coefficients completely by DFT, the present author's group first used LDA and GGA with a surface correction term [37], and later found out that without the surface correction term, the predictions using LDA agreed with experimental better [38–40]. However, for hcp phases, it was found that LDA and GGA results give the lower and upper bounds to the experimental data [41,42], while the results from PBEsol [43] showed much better agreement with experimental data [44]. In predicting the formation energies of compounds in the binary Bi-Nd system, it was found that the results from SCAN show the best agreement with experiments [45]. The SCAN potential was also used in predicting the ground-state configuration of $YNiO_3$ [46].

At the same time, significant processes have also been made in predicting the electronic structures and energetics at finite temperature, including time-dependent DFT (TDDFT) [47–49], random phase approximation (RPA) [50,51], density-matrix functional theory (DMFT) [52–55], DFT+U [56–58], dynamical mean-field theory [59,60], benchmarking with experimental measurements [61], deep neural network machine learning models [62–64], and some other hybrid methods [65]. However, these bottom-up predictions, including the state-of-the-art approaches using effective Hamiltonian, have not resulted in fully satisfactory quantitative agreements with experiments without fitting parameters, as discussed recently by Du et al. [6]. The lack of quantitative agreement between DFT-based predictions and experimental observations is often considered to be related to the approximations in the exchange-correlation functionals.

As mentioned in the introduction, the key information missing in the existing approaches in the DFT community is the lack of ergodicity of configurations. This is in analogy to the symmetry-broken SCAN "where certain strong correlations present as fluctuations in the exact symmetry-unbroken ground-state wavefunction are 'frozen' in symmetry-broken electron densities or spin densities of approximate DFT" as pointed out by Perdew et al. [28]. Those "frozen" symmetry-broken electron densities contribute to the over-all approximated SCAN meta-GGA potentials which deviate from the exact symmetry-unbroken ground-state potentials. By the same token, at finite temperature the system fluctuates among ground-state configuration and symmetry-broken non-ground-state configurations that are derived from the internal degrees of freedom of the ground-state configuration. Those symmetry-broken configurations contribute to the local microscopic structures and can result in the emergent macroscopic structures and behaviors of the system.

It should be mentioned that in calculating the electronic free energy of the ground-state configuration, Kohn and Sham [4] used the finite temperature generalization of ground-state energy of an interacting inhomogeneous electron gas by Mermin [66] and formulated the entropy of thermal electrons at finite temperature. Wang et al [67] added the vibrational contribution and presented the Helmholtz energy for a given stable configuration $k$ as follows

$$F^k = E^{k,0} + F^{k,el} + F^{k,vib} = E^k - TS^k \qquad Eq.\ 1$$

$$E^k = E^{k,0} + E^{k,el} + E^{k,vib} \qquad Eq.\ 2$$

$$S^k = S^{k,el} + S^{k,vib} \qquad Eq.\ 3$$

where $E^{k,0}$ is the static total energy of configuration $k$ at 0K, $F^{k,el}$, $E^{k,el}$, and $S^{k,el}$ are the contributions of thermal electron to Helmholtz energy, total energy, and entropy of configuration $k$ based on the Fermi–Dirac statistics for electrons, and $F^{k,vib}$, $E^{k,vib}$, and $S^{k,vib}$ are the vibrational contributions to Helmholtz energy, total energy, and entropy of configuration $k$ based on the Bose–Einstein statistics for phonons, respectively.

## 3  Zentropy theory: Coarse graining of entropy through integration of DFT and SM

In SM, the partition function of system, $Z$, equals to the sum of the partition functions, $Z^k$, of $m$ configurations as follows

$$Z = \sum_{k=1}^{m} Z^k \qquad Eq.\ 4$$

Conventionally, those configurations can be deduced through basins on energy landscape at finite temperatures in simulations or local structures through scattering in experiments. In principle, the configurations can be tracked all the way down to the pure quantum configurations that do not have any unspecified internal degrees of freedom, thus zero entropy, as discussed in quantum statistical mechanics by Landau and Lifshitz [68]. The partition function of a pure quantum configurations is thus related to its total energy, $E^k$, as follows

$$Z^k = e^{-\frac{E^k}{k_B T}} \qquad \text{Eq. 5}$$

The configurational entropy among pure quantum configurations, $S^{conf}$, and the probability of each configuration, $p^k$, are thus obtained as shown below

$$S^{conf} = -k_B \sum_{k=1}^{m} p^k \ln p^k \qquad \text{Eq. 6}$$

$$p^k = \frac{Z^k}{Z} = e^{-\frac{E^k - F}{k_B T}} \qquad \text{Eq. 7}$$

However, for systems of practical interest, the number of pure quantum states is too large, and their complete sampling is in general intractable. Consequently, there needs to be a set of configurations that are tractable, and their properties can be predicted theoretically without fitting to experimental observations. The current available solution is offered by DFT [3,4], which provides a theoretical framework to coarse-grain the degrees of freedom for electrons and phonons through the Fermi–Dirac and Bose–Einstein statistics, respectively, as shown by Eq. 1 to Eq. 3.

At 0 K, the system experiences only the ground-state configuration. With the increase of temperature, non-ground-state configurations start to appear in the system statistically. Their DFT-predicted entropies contribute to the total entropy of the system. As mentioned in the introduction, this multiscale entropy approach was originally developed for predicting the T-P and T-V phase diagrams of Ce [7,8] and later applied to a number of magnetic and ferroelectric materials as reviewed by the present author [22,69,70] and recently termed as zentropy theory [10,71]. The zentropy theory postulates that the total entropy of the system include both entropies in each configuration and among the configurations, i.e.,

$$S = \sum_{k=1}^{m} p^k S^k + S^{conf} = \sum_{k=1}^{m} p^k S^k - k_B \sum_{k=1}^{m} p^k \ln p^k \qquad Eq.\ 8$$

As schematically shown in Figure 1, the first and second summations in right hand side of Eq. 8 represent the bottom-up and top-down views of the system, respectively.

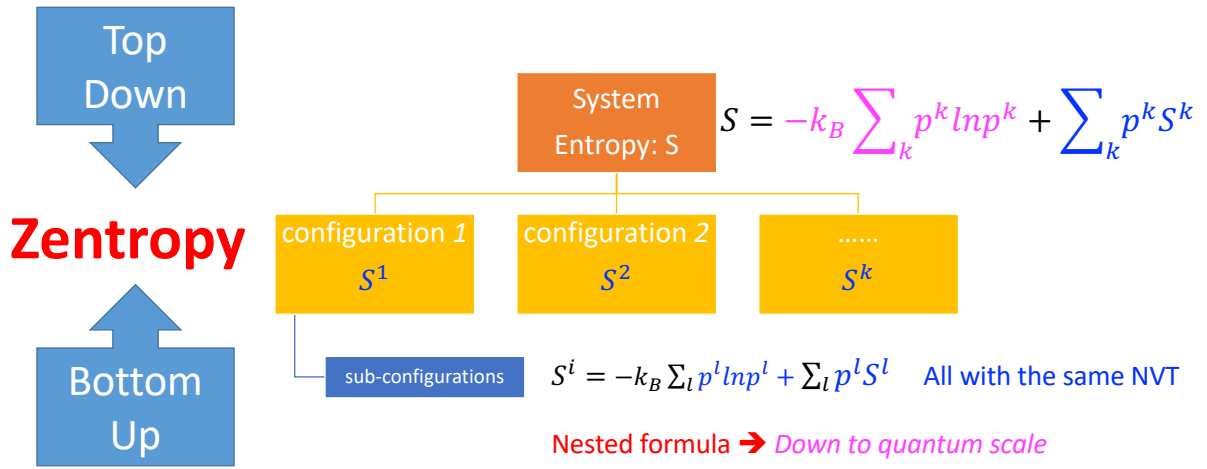

Figure 1: Schematic top-down and bottom-up integration of the zentropy theory [22], Reproduced with permission from CALPHAD 82, 102580 (2023), Copyright 2023 Elsevier.

The Helmholtz energy of the system can thus be obtained as

$$F = \sum_{k=1}^{m} p^k E^k - TS = \sum_{k=1}^{m} p^k F^k - k_B T \sum_{k=1}^{m} p^k \ln p^k \qquad Eq.\ 9$$

Re-arranging Eq. 9 in the form of the partition function, one obtains

$$Z = e^{-\frac{F}{k_BT}} = \sum_{k=1}^{m} e^{-\frac{F^k}{k_BT}} = \sum_{k=1}^{m} Z^k \qquad Eq.\ 10$$

$$p^k = \frac{Z^k}{Z} = e^{-\frac{F^k - F}{k_BT}} \qquad Eq.\ 11$$

The above equations reduce to standard SM when $S^k = 0$, i.e., pure quantum configurations with $F^k = E^k$. With $F^k$ predicted from DFT and $p^k$ from partition functions, the zentropy theory integrates the quantum and statistical mechanics through Eq. 8 to Eq. 11.

Another important outcome of the zentropy theory is the accurate prediction of the free energy landscape including the free energy of unstable states of the system. It was emphasized by Gibbs [5] that the each configuration must be under the same external constraints as the entire system, i.e., no internal relaxations are considered among configurations when they are statistically combined together to form the macroscopically homogenous system as evidenced by Eq. 10. While this macroscopically homogenous system can be in a stable state, it can also be metastable or unstable with respect to internal perturbations. This is significant because the entropy of an unstable state can thus be predicted by the derivative of free energy to temperature. As all individual configurations are stable, the zentropy theory demonstrates that it is the statistical competition among configurations that results in the macroscopic instability of the system at its macroscopic critical point and between inflection points [72]. Immediately passing the instability limit of the system, bifurcation occurs, and the resulted subsystems contain the same configurations as the macroscopically homogenous system though with different statistical probabilities in each subsystem. Consequently, the zentropy theory is capable of predicting the free energy barrier between stable and metastable states, which represent the extrema of anharmonicity and emergent behaviors [73].

It is thus evident that each snapshot of the system in MD or MC simulations represents a statistical mixture of configurations. From the local structures in those snapshots, one may be able to deduce the foundational configurations of the system. One excellent example is $PbTiO_3$ with the ground-state tetragonal ferroelectric configuration and two non-ground-state configurations with head-tail 90° and 180° domain walls, observed experimentally by X-ray absorption fine-structure structure (XAFS) technique [74] and computationally by AIMD simulations [75]. The paraelectric cubic phase at high temperature experimentally observed by X-ray diffractions is a statistical mixture of those ferroelectric configurations that switch among each other much faster than the X-ray's temporal and spatial resolutions, resulting in an apparent cubic structure without macroscopic polarization. Our first attempt to predict the ferroelectric-paraelectric (FE-PE) transition temperature in $PbTiO_3$ with the above three configurations resulted in very encouraging results [11] with ongoing work in the present author's group aiming for more accurate prediction of FE-PE transitions [76].

In principle, this nested formula can be extended to consider more complex systems such as black holes with more complex and degrees of freedom with Eq. 8 denoting one of the sub-systems as recently postulated by the author [22]. It may also help us to understand superconductivity [70], which is actively pursued by the author's group [76].

## 4 Examples of applications of zentropy theory

In recent reviews [22,70], the present author discussed the applications of the zentropy theory in his group and the applications of similar approaches in the literature. One important parameter in DFT-based calculations is the supercell size with a minimum number of atoms required based on the symmetry of a given ground-state configuration. However, for non-ground-state configurations, the supercell size is usually needed to be increased, but often limited by computational resources. For example, in the $PbTiO_3$, the supercell size for the ground-state configuration is five atoms, while for 90° and 180° domain walls, a super cell with 50 atoms is needed in order to obtain the convergence of the domain wall energies [77], which significantly increases the computing expenses in DFT-based phonon calculations.

The author's group has been focusing on magnetic materials in developing the zentropy theory. In magnetic materials the ground-state configurations are either ferromagnetic (FM) or antiferromagnetic (AFM), except non-magnetic (NM) for Ce[7,8], and the non-ground-state configurations are disordered in terms of individual magnetic spins. At high temperatures, they all transform to paramagnetic (PM) phases. The total number of collinear spin configurations is $2^n$ with $n$ being the number of magnetic atoms in supercells used in DFT-based calculations. In our first attempt in 2008 for Ce [7], only NM and FM configurations were considered, requiring only one-atom supercell. A mean-field term commonly used in the literature had to be added to the free energy of system to account for the spin disordering due to the non-ergodicity of two configurations considered. In our next attempt [8], an AFM configuration was added, requiring a two-atom supercell, and the mean-field term was not needed. Both predictions [7,8] showed excellent agreement with available experimental data including the singularity and property divergence at critical point as shown in Figure 2(a) and (b) in terms of its T-P and T-V phase diagrams [8,10]. The positive divergency of thermal expansion at the critical point is predicted, and the glossal thermal expansion in the macroscopical single phase region above the critical point is highlighted by the purple symbols as shown in Figure 2(b). Those experimental observations have not been predicted accurately in the literature previously. It was remarkable that DFT-based calculations of a 2-atom supercell and 3 configurations were able to predict such complex behaviors in Ce using the zentropy theory based on the DFT and SM integration.

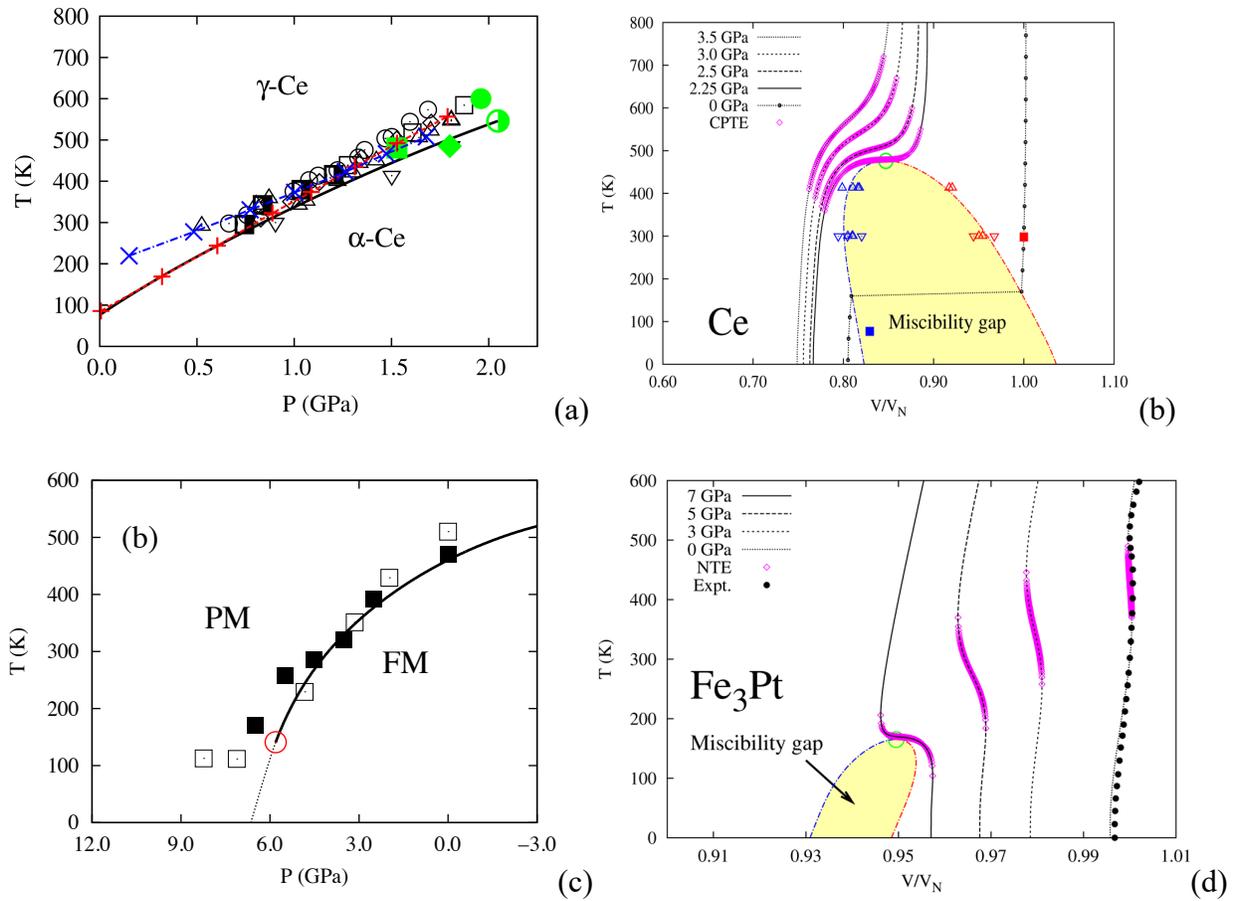

Figure 2: Predicted temperature-pressure and temperature-volume phase diagrams of Ce [8,10] in (a) and (b), and $Fe_3Pt$ [9,10] in (c) and (d) in terms of the zentropy approach. The half-filled circle in (a) and the open red circle in (c) and the green circles in (b) and (d) correspond to the critical

*point, and all other symbols are experimental data. The isobaric volume curves are also plotted in (c) and (d) as a function of temperature. Reproduced with permissions from J. Phys. Condens. Matter 21, 326003 (2009) for (a), Copyright 2009 IOP Publishing, Ltd; from Liu, Z. K., Wang, Y. & Shang, S.-L. Zentropy Theory for Positive and Negative Thermal Expansion, J. Phase Equilibria Diffus. 43, 598–605 (2022) licensed under a Creative Commons license for (b) and (d); and from Philos. Mag. Lett. 90, 851–859 (2010) for (c), Copyright 2010 Taylor & Francis.*

The next magnetic materials that the author's group worked on was $Fe_3Pt$ [9], an INVAR alloy with NTE experimentally observed in a temperature range under ambient pressure. Based on the available computational resources, a 12-atom supercell with 9 Fe atoms was used. This results in $2^9 = 512$ configurations among which 37 configurations are unique. The predicted T-P phase diagram of $Fe_3Pt$ is shown in Figure 2(c) [9,10] with the pressure decreasing from left to right. In its T-V phase diagram shown in Figure 2(d) [9,10], it is observed that the system volume decreases within a temperature range under constant pressure with the experimental measured volumes under ambient pressure superimposed in the figure, showing remarkable agreement between the predictions and measurements.

As in the Ce case, the singularity and property divergence at the critical point in $Fe_3Pt$ are accurately predicted, though with a negative divergency of thermal expansion at the critical point. It is noted the positive and negative slopes of the 2-phase equilibrium lines in Figure 2(a) and (c), respectively, indicating that the high temperature phase with higher entropy has either larger or smaller volume than the low temperature phase with lower entropy. As the high temperature phase has more non-ground-state configurations than the low temperature phase, it was discovered that NTE originates from the non-ground-state configurations with smaller volume than the ground-state configuration [78]. The theories for INVAR in the literature are qualitatively based on phenomenological interpretation of experimental observations [73] and thus not predictive.

Among other magnetic materials that the author's group worked on, $BaFe_2As_2$ stands out as a particularly interesting one [79] in terms of the spin-density-wave (SDW) AFM ordering, a characteristic temperature in terms of a steep decrease of resistivity below approximately 20 – 30K at pressure above 3.0 GPa postulated to be related to its intrinsic superconductivity [80], and as the parent phase of many Fe-based superconductors. In the DFT-based calculations [79], an orthorhombic 40-atom supercell with 16 Fe atom was considered with total $2^{16} = 65536$ configurations, which are too many for brutal force DFT-based calculations. Fortunately, within the temperature range of interest, most of the configurations can be ruled out due to their high energies, and experimental and theoretical investigations in the literatures indicated that the low energy spin configurations are with a stripe-like AFM Fe spin ordering pattern within the *ab* plane and the nearest-neighbor Fe spins parallel along the *c* axis. Consequently, there are 13 such low-energy spin configurations derived from the 40-atom supercell and total 256 with their multiplicities [79]. The calculated 2$^{nd}$-order AFM-PM transition is plotted by the red curve in Figure 3 [79] in remarkable agreement with experimental data plotted in red triangle. Furthermore, the temperature for the change of the thermal probability of the ground-state configuration from 100% to 99.99% as a function of pressure is also plotted in Figure 3, which are in good agreement with the reported characteristic temperature [80]. At lower pressure, such characteristic behavior was not observed. This is an interesting topic related to superconductivity discussed in the literature [81] and as part of the on-going research in the author's group [70,76].

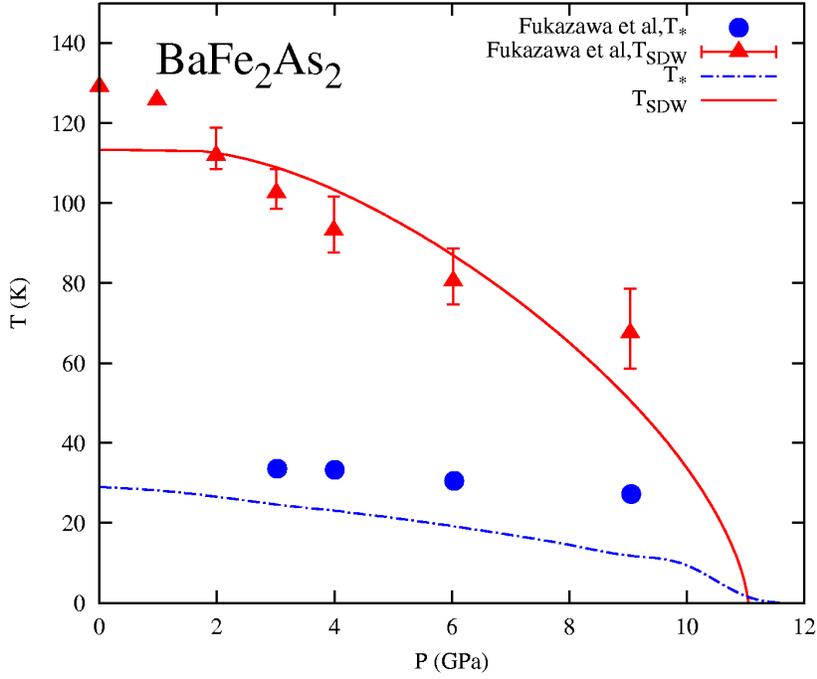

Figure 3: Predicted SDW ordering temperature ($T_{SDW}$, red curve) and characteristic temperature ($T_*$, blue curve) plotted with respect to pressure with experimental data (symbols [80]) in orthorhombic-$BaFe_2As_2$ [79]. Reproduced with permission from Int. J. Quantum Chem. 111, 3565–3570 (2011), Copyright 2011 John Wiley and Sons.

More recently, the author's group applied the zentropy theory to $YNiO_3$ with the so-called strongly correlated physics [6,46]. The ground-state configuration of $YNiO_3$ is with the S-type AFM $P2_1/n$ structure with a primitive unitcell of 20 atoms [46]. It was revealed that half of Ni atoms are with null spin moment, and a 1×2×2 supercell with respect to the 20-atom primitive unitcell and total 80 atoms was then created to enumerate its $2^8 = 256$ magnetic spin configurations with 37 being unique. With their Helmholtz energies obtained from the DFT-based phonon calculations, the predicted T-P phase diagram is shown in Figure 4(a) with all the AFM-PM transitions being 2$^{nd}$-order thus without critical point in the system [6]. The predicted magnetic transition temperature is 144 K, one degree below the experimentally measured temperature of 145 K. However, if $E^k$ is used for the partition function of each configuration, the predicted transition temperature is 81 K as marked in Figure 4(a), demonstrating the importance to consider the entropy of individual configurations in their partition functions as part of the zentropy theory. Furthermore, an approach was developed to evaluate the short-range ordering (SRO) of magnetic spin in terms of the standard deviation of the distribution of total magnetic moments around Ni atom within its 1$^{st}$ coordination sphere, computed by the probability of various configurations, with respect to the same standard deviation in the special quasi-random structure (SQS) spin model that mimics the fully random distribution of spins. Thus obtained SRO for YNiO$_3$ is plotted in Figure 4(b).

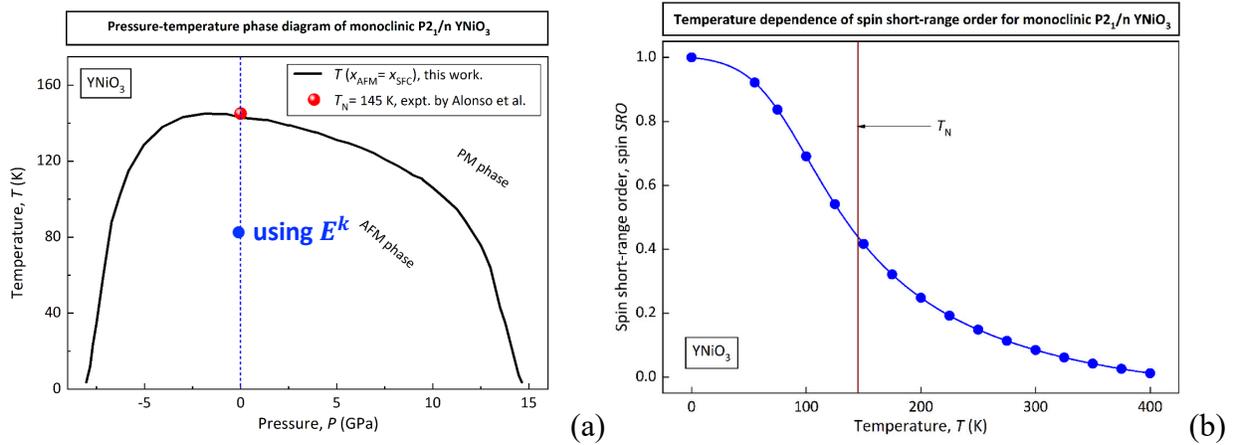

Figure 4: (a) Predicted temperature-pressure phase diagram of YNiO$_3$ (sloid curve, 2$^{nd}$-order magnetic phase transition using $F^k$ for each configuration) separating the AFM and PM phase regions with the experimental (red dot) and predicted (blue dot, using $E^k$ for each configuration)

*superimposed [6]; and (b) Predicted short-rang magnetic spin ordering in $YNiO_3$ [6]. Reproduced with permission from Mater. Today Phys. 27, 100805 (2022), Copyright 2022 Elsevier.*

In Section 3, it was commented that the zentropy theory can predict the free energy landscape. This is reflected in the predicted T-P and T-V phase diagrams of Ce and $Fe_3Pt$ shown in Figure 2. In both systems, the critical point separates the $1^{st}$- and $2^{nd}$-order transitions. The Helmholtz energy of a system as a function of volume is obtained by Eq. 10 through the summation of the partition functions of all configurations with the same temperature and volume. In the $1^{st}$-order transition region, the Helmholtz energy thus obtained has two basins and an apex between them at each temperature. The minimization of the Helmholtz energy with respect to the formation of two phases with different probabilities of various configurations results in the T-V phase diagrams shown in Figure 2(b) and (d), i.e., the regions marked by miscibility gap with respect to volume. Between the apex and a basin, there is an inflection point, representing the limit of the stability/instability of the system, while the height of the apex denotes the transformation Helmholtz energy barrier between the two basins.

The $2^{nd}$-order transition is thermodynamically defined by continuity in $1^{st}$ derivatives and discontinuity in $2^{nd}$ derivatives of free energy to potentials such as temperature and pressure (see e.g., Table 1 in the author's recent review [22]). Due to the limited supercell sizes used in DFT-based calculations, those derivatives are not completely discontinuous, but with a bump on magnetic heat capacity due to spin disordering. On the other hand, it was found that when the critical point is approached from the temperature below, the thermal probability of the ground-state configuration reaches 50% at the critical point and decreases further at higher temperature.

This could be understood that the ground-state configuration loses percolation when its probability decreases to below 50%. This criterion was used to define the 2$^{nd}$-order transition in Figure 2(a) and (c), which is higher than the temperatures corresponding to the maximum magnetic heat capacity. It was recently shown in the literature that the temperature for the maximum magnetic heat capacity is systematically lower than that from the maximum total heat capacity [82].

In magnetic materials studied in our publications so far, collinear spin configurations have been used. While noncollinear configurations have been suggested in the literature, however, most of them seem unstable at 0K, indicating that the system may not be able to embrace them with any residence time, but passes through them instantaneously when the system fluctuates between configurations. Furthermore, the applications of the zentropy theory have demonstrated that the strongly correlated physics are due to both the spin interactions within each configuration and the statistical competition among configurations with the latter being highly nonlinear as shown by Eq. 8. It can thus be concluded that only the combination of both terms in Eq. 8 can account for the total entropy and the extreme of anharmonicity in a system even though each configuration is well described by the quasiharmonic approximation. This is very much in analogy to the parable of the blind men and the elephant discussed by Perdew et al. [28] in connection with the discussion of perspectives on broken symmetry and strong correlation in many-electron systems.

## 5    Summary

In the present paper, the zentropy theory for accurate prediction of Helmholtz energy as a function of temperature and volume through integration of DFT and statistical mechanics is discussed. The zentropy theory postulates that the entropy of a system contains contributions from the entropies of the ground-state and symmetry-breaking non-ground-state configurations of the system and the configurational entropy among the configurations. Consequently, the partition function of each configuration needs to be calculated from its Helmholtz energy instead of total energy commonly used in the literature. It is demonstrated that the zentropy theory can accurately predict the both $1^{st}$- and $2^{nd}$-order magnetic transitions and the Helmholtz energy barriers in $1^{st}$-order magnetic transitions with the Helmholtz energies of their individual spin configurations predicted from DFT-based calculations using the potentials developed by Perdew that capture the symmetry-breaking configurations at the electronic scale. It is emphasized that the properties predicted by the DFT-based calculations of the ground-state configuration alone are intrinsically different from experimental observations at finite temperature that contain contributions from symmetry-breaking non-ground-state configurations. The preset author believes that the zentropy theory can be applied to predict other emergent behaviors provided both ground-state and stable symmetry-breaking non-ground-state configurations of the system can be defined. It is articulated that the anharmonicity and its extreme at a critical point primarily come from the statistical competition among ground-state and non-ground-state configurations.

**Acknowledgements.** The present review article covers research outcomes supported by multiple funding agencies over many years with the most recent ones including the Endowed Dorothy Pate Enright Professorship at the Pennsylvania State University, U.S. Department of Energy (DOE) Grant No. DE-SC0023185, DE-AR0001435, DE-NE0008945, and DE-NE0009288, and U.S. National Science Foundation (NSF) Grant No. NSF-2229690. The author thanks John Perdew for explaining their recent works on symmetry breaking due to fluctuations of various wavevectors [28,29], his interest in the present work, and for the opportunity to write the present manuscript for this special John Perdew Festschrift issue.